\documentclass[10pt]{article}
\usepackage{ulem}
\usepackage{color}
\usepackage{epsfig}
\usepackage{graphicx}
\usepackage{subfigure}
\usepackage{ulem}
\usepackage{amsfonts}
\usepackage{epsfig,bm}
\usepackage{graphicx}
\usepackage{comment}
\newcommand{\be}{\begin{equation}}
\newcommand{\ee}{\end{equation}}
\newcommand{\bea}{\begin{eqnarray}}
\newcommand{\eea}{\end{eqnarray}}
\newcommand{\bean}{\begin{eqnarray*}}
\newcommand{\eean}{\end{eqnarray*}}

\begin{document}
\begin{titlepage}
\title{On Intrinsic Geometric Stability of Controller}
\author{}
\date{Stefano Bellucci $^{a}$ \thanks{\noindent bellucci@lnf.infn.it}, \
 Bhupendra Nath Tiwari $^{a}$ \thanks{\noindent bntiwari.iitk@gmail.com},\
N. Amuthan $^{b,}$ \thanks{\noindent amuthannadar@gmail.com} \ and
S. Krishnakumar$^{c,}$ \thanks{\noindent sabakrish@gmail.com} \\
\vspace{0.5cm}
$^a$INFN, Laboratori Nazionali di Frascati, Via E. Fermi 40, 00044 Frascati, Italy.\\
$^b$ PET Engineering College, Department of Electrical and
Electronic Engineering, Valliyoor, Tamil Nadu, India\\
$^c$ Department of Physical Science, Faculty of Applied Sciences,
Vavuniya Campus of the University of Jaffna, Kurumankadu,
Vavuniya, Sri Lanka}
 \maketitle \abstract{This work explores the role of the intrinsic fluctuations in
finite parameter controller configurations characterizing an
ensemble of arbitrary irregular filter circuits. Our analysis
illustrates that the parametric intrinsic geometric description
exhibits a set of exact pair correction functions and global
correlation volume with and without the variation of the mismatch
factor. The present consideration shows that the canonical
fluctuations can precisely be depicted without any approximation.
The intrinsic geometric notion offers a clear picture of the
fluctuating controllers, which as the limit of the ensemble
averaging reduce to the specified controller. For the constant
mismatch factor controllers, the Gaussian fluctuations over
equilibrium basis accomplish a well-defined, non-degenerate, flat
regular intrinsic Riemannian surface. An explicit computation
further demonstrates that the underlying power correlations
involve ordinary summations, even if we consider the variable
mismatch factor controllers. Our intrinsic geometric framework
describes a definite character to the canonical power fluctuations
of the controllers and constitutes a stable design strategy for
the parameters.}
%$D$-brane charge= electric charge
\vspace{1.5cm}

\textbf{Keywords: Correlation, Fluctuation, Geometry, Controller, Stability} \\

\end{titlepage}

\section{Introduction}

Stable design of controllers is one of the most interesting
research issue since the proposition of the robust controllers.
Such a principle has been extremely successfully which applies in
both the domestic and industrial application \cite{HN}. This
follows because of the fact that it is easy in implementation and
requires less number of parameter tuning. Further, the stably
designed controllers are the best alternative to the existing
controllers. This follows from the fact that the intrinsic
geometric design takes an account of the model uncertainties and
specifically allows for a clear-cut determination of the
controller settings and parameters. Such an observation is
supported from the fact that a class of controllers remains
insensitive to the time delay and parametric deviation for the
first order systems \cite{HN}.

Global Stability phenomenon are the subject matter of \cite{1}.
Also, the parametric approach robust controllers have been brought
out in the picture with the notions of \cite{2}. An important
filter design taking an account of the load disturbance is
proposed \cite{3} in order to improve the speed of the tuning
response function. The stability of such a class of controllers
depends only on the domain of the controller parameters and so the
construction of a nominal plant. In addition, although the system
can reached control input saturation, but the stability of the
designed controller can be maintain to only depend on the
parameters of the intrinsically designed controller and that of
the plant, as per the outlines of the Ref. \cite{4}. Similar
direct model reference adaptive controllers \cite{5,6,8} are
explored in diverse applications.

Parametric model controller design is one of the best methods to
improve the robust recital of the controller because of the fact
that a polynomial approach is involved in the performance
improvement of the controller. The parameters concerning the speed
of the controller depend on the model parameters and mismatch
factor of the low pass filter circuit. This is because the
robustness of the controllers and their performance depend on the
model parameters $\{a, b \}$ and the mismatch factor $f$ of the
filter. In the non-linear domain of the above parameters, our
intrinsic geometric analysis provides a stable characterization of
the controllers. From the viewpoint of the present interest,
Ref.\cite{9} leads us to provide an appropriate mathematical
design and parameterization for the controllers. The parameterized
equation controller block diagram is further brought out into the
present attention. From the out-set of the above reference
\cite{9}, we can track the desired trajectory and minimize the
plant error. The key issues concerning the controllers are the
speed response, low pass filter and geometric model. The robust
performance of such a controller is based on these parameters
\cite{10, 11}.

A novel approach is thus made possible in the history of
controllers via the present investigation. The method as outlined
above is very general in its own and it leaves different possible
versions to be explored further. One of the key issues is an
appropriate design of the controller circuits. To design the
parametric internal model, one traditionally assumes that one of
the parameters of the controller is in the subset of parametric
family of the controller. It is sometime called finite dimensional
parametric model. The filtering operations in these controllers
are associated with the parameter involved. The distribution of
parameters in the parametric model can be taken to be finite
dimensional. Furthermore, the model reference parameterizations
are taken into an account \cite{12} and the associated
parameterizable methods are used to obtain the parametric
controllers.

Based on the conventional design method, we offer intrinsic
stability analysis of the controllers. The linear parametric
polynomial approach which improves structures of the bounded
parametric controller integral-derivative (id) designs, is
explicitly presented. Improvements in the limiting linear
parametric controllers are shown. It turns out that the present
intrinsic geometric notion is particularly well suited for the
practical applications. The mathematical design procedure thus
taken can be extended for any controllers. In fact, we have a
clear picture of further investigation. This has been a real
bestow for stabilization of the conventional controllers. The
present analysis of controller can further be used to explore the
intrinsic nature of the unmodeled part of the plant and the
associated bounded disturbances arising from the fluctuations of
parameters and mismatch factor.

Intrinsic geometric modelings involving equilibrium configurations
of the extremal and the non-extremal black holes in string theory
\cite{9601029v2,9411187v3,9504147v2,
0409148v2,9707203v1,0507014v1, 0502157v4,0505122v2} and $M$-theory
\cite{0209114,0401129,0408106,0408122} possess rich intrinsic
geometric structures \cite{0606084v1,SST,bnt, BNTBull, BNTBull08}.
There has been much well focused attention on the equilibrium
perspective of black holes, and thereby explicates the nature of
concerned parametric pair correlations and associated stability of
the solutions containing a large number of branes and antibranes.
Besides several general notions which have earlier been analyzed
in the condensed matter physics \cite{RuppeinerRMP,
RuppeinerA20,RuppeinerPRL,RuppeinerA27,RuppeinerA41,RuppeinerPRD78},
we consider specific controller configurations thus mentioned with
equilibrium parameters and analyze possible parametric pair
correlation functions and their correlation relations. Basically,
the investigation entails an intriguing feature of the underlying
fluctuations which are defined in terms of the parameters.

Given a definite covariant intrinsic geometric description of a
consistent controller configuration, one can expose (i) for what
conditions the considered configuration is stable?, (ii) how its
parametric correlation functions scale in terms of a set of chosen
fluctuating circuit parameters? In this process, one can enlist a
complete set of non-trivial parametric correlation functions of
the controller configurations \cite{9}. It may further be
envisaged that similar considerations remain valid over the black
hole solutions in general relativity
\cite{gr-qc/0601119v1,gr-qc/0512035v1,gr-qc/0304015v1, 0510139v3},
attractor black holes \cite{9508072v3,9602111v3,new1,new2,
0702019v1,0805.1310} and Legendre transformed finite parameter
chemical configurations \cite{Weinhold1, Weinhold2}, quantum field
theory and the associated Hot QCD backgrounds \cite{BNTSBVC}.
Thus, the differential geometry plays an important role in the
thermodynamic study of the controllers.

In this paper, we analyze the stability of the controllers under
the fluctuation of the parameters and the mismatch factor. The
controller under consideration is depicted in the Fig.[1]. The
stability is demonstrated for a suitable design and its
parameterizations. From the general parametrization equation of
the controller, the stability of block diagram is drawn into
attention. To the best of authors' knowledge, this approach is
made possible for the first time towards the intrinsic geometric
stability analysis of controllers. The proposed method is very
general and different Legendre transformed versions of the present
analysis are possible. By employing the standard notion of the
intrinsic Riemannian geometry, the rest of the sections are
devoted to the local and global stability properties of
fluctuating controllers.
\section{Fluctuations in the Controllers}

The controller circuit of the present interest is depicted in the
Fig.[1]. Although the analysis of the present investigation
remains for any controllers, nevertheless we illustrate it for a
class of controllers, which are of an immediate interest as shown
in the Fig.[1]. Here, $s$ is a complex signal having a modulus and
an angle of phase. In the subsequent analysis, we denote a locally
constant signal by the corresponding uppercase notation $S$.
Furthermore, we show that the investigation of the stability
analysis is valid for the general controllers, and demonstrate
that our approach remains consistent with the other existing ones.
Given the controller, we consider the intrinsic geometric
stability of the underlying low pass filter system. The parametric
stability criterion offers a proficient method to determine the
parameters of the circuit and thereby to design the controller as
per ones requirement.

Ref \cite{Gilglio} implements the principle for as associated
class of controllers. From a close perspective, such an analysis
of the controllers takes an account of the model uncertainties.
Specifically, it allows a straightforward relation of controller
settings with the associated model parameters. Notice further that
the first order consideration of the controllers is insensitive to
time delay and parameter deviation, and the output is
approximately equal to the PI controllers \cite{HN}. The response
of the controller is sluggish, although it does not have important
overshoot effects \cite{IS} and the integral action of such a
controller is used to eliminate the offset of the system. To
explore the stability of the first order controllers \cite{HOS},
we design stable controllers from the perspective of the intrinsic
geometry.
%Before proceeding further, it is worth
%mentioning the following facts regarding the IMCs:
%\begin{itemize}
%\item The closed loop transfer function of IMC is given by
%\begin{scriptsize}
%\begin{eqnarray}
%Y(S)= \frac{G_{imc}(S)G_p(S)R(S)+[1-G_{imc}(S)G_{inv}(S)]d(S)}
%{1+[G_p(S)-G_{inv}(S)]G_{imc}(S)}
%\end{eqnarray}
%\end{scriptsize}
%\item First order plant model
%\begin{scriptsize}
%\begin{eqnarray}
%G_p= K_p \frac{S-a}{S+b}
%\end{eqnarray}
%\end{scriptsize}
%\item Internal model
%\begin{scriptsize}
%\begin{eqnarray}
%G_{inv}= \frac{S+b}{S-a}
%\end{eqnarray}
%\end{scriptsize}
%\item Low pass filter F is used to avoid model mismatch
%\begin{scriptsize}
%\begin{eqnarray} F= (1+fS)^{-n}
%\end{eqnarray}
%\end{scriptsize}
%\item The speed response tuning parameter $f$ and order of filter $n$ satisfy
%\begin{scriptsize}
%\begin{eqnarray}
%H=FG_{inv}= \frac{(S+b)}{(1+fS)^{n}(S-a)}
%\end{eqnarray}
%\end{scriptsize}
%\end{itemize}

Having mentioned the domains of the parameters, we consider the
following two specification of the controllers, \textit{viz.},
constant mismatch factor controllers, and variable mismatch factor
controllers. For arbitrary n$^{th}$ order low pass filter, the
controllers of the intrinsic interest are
\begin{scriptsize}
\begin{eqnarray} \label{controller}
G_{con}(a,b,f):=\frac{(S-a)(S+b)}{(1+fS)^n(S-a)^2-(S+b)^2}
\end{eqnarray}
\end{scriptsize}
Notice further that the aim of the present paper is to expose the
power of the intrinsic geometry. In this concern, the controller
described by the Eqn.(\ref{controller}) serves only as an example
of the present consideration. Subsequently, our analysis  as the
exposition of the intrinsic geometric investigation remains valid
for any smooth controller and thus the above class of the
controllers. In order to begin the subsequent intrinsic geometric
analysis of fluctuations, we introduce the correlation in the
controller arising from the fluctuations of the circuits
parameters. Thus, we consider an ensemble of controllers
fluctuating over the limiting Gaussian ensemble. In this analysis,
we consider that the controllers can have non-zero fluctuations
due to the vibrations of frequencies, residual ripple factors in
the filter circuits, and possible other practical uses. This
follows from the fact that we do not restrict ourselves in the
specific domains of filter circuit used in the controller.

Consequently, we allow an ensemble of limiting configurations with
finite fluctuations in an arbitrary non linear domain of the
parameters and thereby analyze the nature of a class of generic
controllers. We also take an account of the variable mismatch
factor defining the speed response tuning parameter of the
controller. Notice further that the analysis of the present
exposition is valid for all range of the parameters of the
controllers. Physically, their deviation from the origin signifies
a contribution of the non-linear effects of the controller.
Specifically, the values $a=0$ and $b=0$ of the parameter signify
a purely linear model controller. Subsequently, the stability of
the controller can be analyzed in the non large frequency domains.
This is because the present analysis is devoted for the
controllers which are defined as the $n^{th}$-order low pass
filter circuit. Such a notion is observed, when there are ripples
in the filter circuits of the controller.
\begin{figure}
\hspace*{0.5cm}
\includegraphics[width=8.0cm,angle=0]{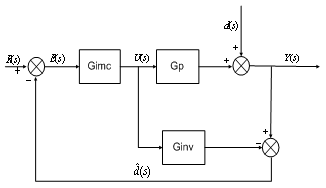}
\caption{A class of controllers as the function of parameters $a$
and $b$, describing fluctuations of the controller with a given
mismatch factor $f$.} \vspace*{0.5cm}
\end{figure}
\section{Constant Mismatch Controllers}
Let us first describe the intrinsic stability of the controller
with a given mismatch factor. The correlations are described by
the Hessian matrix of the controller, defined with a set of
desired corrections over a chosen model mismatch factor $f$ under
the tuning response function. Following Eqn.(\ref{controller}),
the components of the metric tensor defined as
$Hess(G_{con}(a,b))$ reduce to the following expressions:
\begin{scriptsize}
\begin{eqnarray} \label{corconst}
g_{aa} &=& \frac{ 2(S+b)(S-a)(1+fS)^n
\{3(S-a)^2(1+fS)^{n}+(S+b)^2\}}
{\{(1+fS)^n(S-a)^2-(S-b)^2\}^{3}} \nonumber \\
g_{ab} &=& \frac{\{(S-a)^4(1+fS)^{2n}
+6(S-a)^2(S+b)^2(1+fS)^n+(S+b)^4\} }
{\{(1+fS)^n(S-a)^2-(S-b)^2\}^{3}} \nonumber \\
g_{bb} &=& \frac{2(S+b)(S-a)\{3(1+fS)^n(S-a)^2+(S+b)^2\} }
{\{(1+fS)^n(S-a)^2-(S-b)^2\}^{3}}
\end{eqnarray}
\end{scriptsize}
In this framework, we observe that the geometric nature of
parametric pair correlations divulges the notion of fluctuating
controllers. Thus, the fluctuating controllers may be easily
determined in terms of the intrinsic parameters of the underlying
circuit configurations. Moreover, it is evident for a given
controller that the principle components of the metric tensor
signify self pair correlations, which are positive definite
functions over a range of the parameters. Physically, this
signifies a set of heat capacities against the intrinsic
interactions on the configuration $(M_2(R),g)$ of the controller.

It is worth mentioning that the controllers turn out to be
well-behaved for the generic values of the parameters. Over the
domain of the circuit parameters $\{ a,b \}$, we notice that the
Gaussian correlations form stable correlations, if the determinant
of the metric tensor
\begin{scriptsize}
\begin{eqnarray}
Det(g)&=& -\frac{\{(1+fS)^{n}(S-a)^2+(S+b)^2\}^2}
{\{(1+fS)^n(S-a)^2-(S+b)^2\}^{4}}
\end{eqnarray}
\end{scriptsize}
remains a positive function on the parametric surface
$(M_2(R),g)$. What follows further is that we specialize ourselves
for the physical values of the parameters, and subsequently, we
analyze the stability for $a=0, b=0$ corresponding to the linear
controllers Fig.[2].
\begin{figure}
\vspace*{-0.5cm}
\includegraphics[width=8.0cm,angle=0]{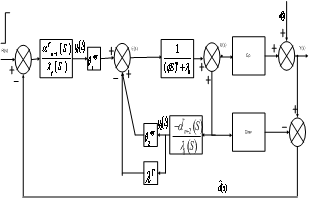}
\caption{Linear controller as the limiting values of the
parameters $a=0$ and $b=0$, described with an intrinsic mismatch
factor $f$.} \vspace*{-0.2cm}
\end{figure}
Under such a limiting specification of the parameters, the local
correlation functions reduce to the set of following expressions:
\begin{scriptsize}
\begin{eqnarray}\label{limitingconstmismatchcorr}
g_{aa} &=&  \frac{6}{S^4}
\frac{(1+fS)^{2n}+2(1+fS)^n}{((1+fS)^n-1)^3}, \ \
g_{bb} = \frac{2}{S^4} \frac{(3(1+fS)^n+1)}{((1+fS)^n-1)^3}  \nonumber \\
g_{ab} &=& \frac{1}{S^2} \frac{(1+fS)^{2n}+6(1+fS)^n+1}{((1+fS)^n-1)^3} \nonumber \\
\end{eqnarray}
\end{scriptsize}
For $a=0, b=0$, the determinants of the metric tensor reduce to
the following expression:
\begin{scriptsize}
\begin{eqnarray}
Det(g)= -\frac{1}{S^4}
\frac{(1+fS)^{2n}+2(1+fS)^n+1}{((1+fS)^n-1)^4}
\end{eqnarray}
\end{scriptsize}
The behavior of the determinant of the metric tensor shows that
such controllers become unstable for specific values of the
mismatch factor. For $S=1$ and $n=1$, the nature of the
determinant of the metric tensor is depicted in the Fig.[3]. It is
worth mentioning further that the constant mismatch controllers
become highly unstable in the limit of vanishing mismatch factor.
Specifically, the system acquires a throat in the regime of $\vert
f \vert \leq 0.7$.
\begin{figure}
\hspace*{0.5cm}
\includegraphics[width=8.0cm,angle=0]{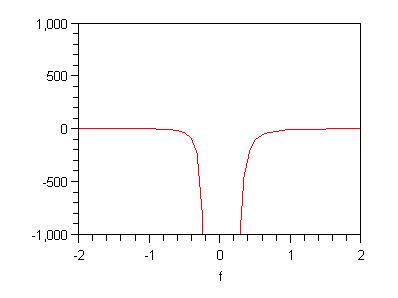}
\caption{The determinant of the metric tensor plotted as the
function of an intrinsic mismatch factor $f$ with given parameters
$a$ and $b$, describing the fluctuations in the controllers.}
\vspace*{0.5cm}
\end{figure}

In order to explain the nature of transformation of the parameters
$\{a,b\}$ forming an intrinsic surface, we examine the functional
behavior of the associated Christoffel connections. A direct
computation yields that the limiting non-trivial Christoffel
connections reduce to the following expressions:
\begin{scriptsize}
\begin{eqnarray}
\Gamma_{aaa} &=& \frac{3}{S^3} \frac{(1+fS)^{3n}+6(1+fS)^{2n}+(1+fS)^n}{((1+fS)^n-1)^4} \nonumber \\
\Gamma_{aab} &=&  \frac{1}{S^3} \frac{(1+fS)^{3n}+14(1+fS)^{2n}+9(1+fS)^n}{((1+fS)^n-1)^4} \nonumber \\
\Gamma_{aba} &=& \frac{1}{S^3} \frac{(1+fS)^{3n}+14(1+fS)^{2n}+9(1+fS)^n}{((1+fS)^n-1)^4} \nonumber \\
\Gamma_{abb}&=& \frac{1}{S^3} \frac{9(1+fS)^{2n}+14(1+fS)^n+1}{((1+fS)^n-1)^4} \nonumber \\
\Gamma_{bba} &=&  \frac{1}{S^3} \frac{9(1+fS)^{2n}+14(1+fS)^n+1}{(1+fS)^n-1)^4} \nonumber \\
\Gamma_{bbb} &=& \frac{3}{S^3}
\frac{(1+fS)^{2n}+6(1+fS)^n+1}{((1+fS)^n-1)^4}
\end{eqnarray}
\end{scriptsize}

The present investigation shows that a typical controller is
globally un-correlated over all possible Gaussian fluctuations of
the parameters $\{a,b\}$. As the matter of fact, the correlation
length of underlying nearly equilibrium system is global
characterized by the scalar curvature of $(M_2,g)$. This follows
from the fact that the scalar curvature, arising from the
definition of the Gaussian fluctuations over the parameters $\{a,
b\}$, vanishes identically for the constant mismatch factor
controllers.

For the above type of controllers with a chosen mismatch factor,
we see that the Riemann Christoffel curvature tensor vanishes
identically over the entire $\{a,b\}$ surface. Thus, the present
intrinsic geometric analysis anticipate that a constant mismatch
controller is always a non-interacting system over the surface of
fluctuating parameters $\{a,b\}$.
\section{Fluctuating Mismatch Controllers}
In the present section, we explore the nature of an ensemble of
generic controllers generated with a variable mismatch factor. To
consider the most general case, we chose the controller as the
function of mismatch factor along with the other system
parameters. When the mismatch factor is allowed to fluctuate, we
exploit the definition of the Hessian function
$Hess(G_{con}(a,b,f))$. Following Eqn.(\ref{controller}), we see
for the variable mismatch factor controllers have a set of
interesting properties. It follows that the pure pair correlations
$\{ g_{aa}, g_{ab}, g_{bb} \}$ between the parameters $\{a, b\}$
remain the same as depicted in the Eqn.(\ref{corconst}) for the
constant mismatch factor controllers. The remaining parametric
pair correlations, involving the variation of the mismatch factor
$f$, are given by the following set of equations:
\begin{scriptsize}
\begin{eqnarray}
%g_{aa} &=& \frac{2(S+b)(S-a) (1+f S)^n \{3(1+f S)^{n}
%(S-a)^2+(S+b)^2\} }
%{\{(1+f S)^n (S-a)^2-(S+b)^2\}^{3}}  \nonumber \\
%g_{ab} &=& \frac{(1+f S)^{2 n} (S-a)^4+6(1+f S)^n(S-a)^2(S+b)^2+(S+b)^4}
%{\{(1+f S)^n (S-a)^2-(S+b)^2\}^{3}} \nonumber \\
g_{af} &=& -\frac{ nS(S+b)(S-a)^2 (1+f S)^{n-1} \{(1+f S)^{n}
(S-a)^2+3(S+b)^2\}}
{\{(1+f S)^n (S-a)^2-(S+b)^2\}^{3}} \nonumber \\
%g_{bb} &=& \frac{2 (S+b) (S-a) \{3 (1+f S)^n (S-a)^2+(S+b)^2\}}
%{\{(1+f S)^n (S-a)^2-(S+b)^2\}^{3}} \nonumber \\
g_{bf} &=& -\frac{ n S(S-a)^3 (1+f S)^{n-1} \{(1+f S)^{n}
(S-a)^2+3 (S+b)^2\}}
{\{(1+f S)^n (S-a)^2-(S+b)^2\}^{3}} \nonumber \\
g_{ff} &=& \frac{ n S^2(S-a)^3 (S+b)(1+f S)^{n-2} \{(n+1)(1+f
S)^{n}(S-a)^2+(n-1) (S+b)^2\}} {\{(1+f S)^n (S-a)^2-(S+b)^2\}^{3}}
\end{eqnarray}
\end{scriptsize}
We see that the fluctuations of the mismatch factor controller
comply physically expected conclusions. In particular, the heat
capacities, defined as the self-pair correlations, remain positive
quantities for well-defined controllers. A straightforward
computation further demonstrate the overall nature of the
parametric fluctuations. A variable mismatch factor controller is
stable under the set of Gaussian fluctuations, if the associated
principle minors $\{p_2, p_3 \}$ remain positive functions on the
manifold $(M_3,g)$. Subsequently, an explicit computation shows
that the stability constraint on the $ab$-surface is given by
\begin{scriptsize}
\begin{eqnarray}
p_2:=-\frac{m_0+m_1(1+f S)^{2 n}+m_2 (1+f S)^{4 n}} {((1+f S)^n
(S-a)^2-(S-b)^2)^6}
\end{eqnarray}
\end{scriptsize}
where the polynomials $m_i(S)$ are given as
\begin{scriptsize}
\begin{eqnarray}
m_0(S)&=& S^8+8 b S^7+28 b^2 S^6+56 b^3 S^5 +70 b^4 S^4 +56 b^5 S^3+28 b^6 S^2 +8 b^7 S+b^8  \nonumber \\
m_1(S)&=& -2 S^8 +8(a-b)S^7-4 (3b^2+3a^2-8ab) S^6+8 (a^3-b^3+ 6a
b^2-6a^2 b)S^5  \nonumber \\ && -2 (a^4+b^4+36a^2b^2-16 a^3 b-16
b^3 a) S^4+8( 6 a^3 b^2-6 b^3 a^2- a^4 b+ b^4 a) S^3  \nonumber \\
&&
-4(3a^4 b^2+3 b^4 a^2-8 b^3 a^3)S^2-8 b^3 a^3 (a-b) S-2 b^4 a^4  \nonumber \\
m_2(S)&=& S^8-8 a S^7+28 a^2 S^6-56 a^3 S^5+70 a^4 S^4 -56 a^5
S^3+28 a^6 S^2-8 a^7 S+ a^8
\end{eqnarray}
\end{scriptsize}
The stability constraint on the entire configuration is determined
by the determinant of the metric tensor
\begin{scriptsize}
\begin{eqnarray} \label{gmismatch}
Det(g)&=& -\frac{(S-a)^3 (S+b) n S^2}{( 1+f S)^{2}((1+f S)^n
(S-a)^2-(S+b)^2)^{7}} g_1(a,b,f)
\end{eqnarray}
\end{scriptsize}
Notice that the co-ordinate charts on $(M_3,g)$ are described by
the parameters $\{a,b\}$ and mismatch factor $f$ of the
controller. In Enq(\ref{gmismatch}), the determinant of the metric
tensor can have a positive value, if the functions $ g_1(a,b,f)$
defined as
\begin{scriptsize}
\begin{eqnarray} \label{g1abf}
g_1(a,b,f)&:=& (n-1) h_1(S)(1+f S)^n +(3n-1) h_2(S) (1+f S)^{2 n}
\nonumber \\&& +(3n+1) h_3(S)(1+f S)^{3 n}+(n+1) h_4(S) (1+f S)^{4
n}
\end{eqnarray}
\end{scriptsize}
take a negative value over $(M_3,g)$. In Eqn.(\ref{g1abf}), it
turns out that $\{ h_k(S)\}$ reduce to the following polynomials:
\begin{scriptsize}
\begin{eqnarray}
h_1(S)&=& S^6 +20 S^3 b^3+6 S^5 b+15 S^2 b^4+15 S^4 b^2+6 b^5 S+ b^6 \nonumber \\ % (s+b)^6
h_2(S)&=& S^6-2 S^5 a+4 S^5 b-8 S^4 ab+6 S^4 b^2+S^4 a^2+4 S^3 a^2
b-12 S^3 a b^2 \nonumber \\ &&
+4 S^3 b^3+6 S^2 a^2 b^2+S^2 b^4-8 S^2 a b^3+4 S a^2 b^3-2 S a b^4+a^2 b^4 \nonumber \\
h_3(S)&=& S^6-4 a S^5+2 S^5 b+6 a^2 S^4+S^4 b^2-8 a S^4 b-4 a^3
S^3 +12 a^2 S^3 b \nonumber \\&&
-4 a S^3 b^2+a^4 S^2-8 a^3 S^2 b+6 a^2 S^2 b^2-4 a^3 S b^2+2 a^4 S b+a^4 b^2 \nonumber \\
h_4(S)&=& S^6-6 a S^5+15 a^2 S^4-20 a^3 S^3+15 a^4 n S^2-6 S a^5+a^6 %(s-a)^6
\end{eqnarray}
\end{scriptsize}
It is not difficult to compute an exact expression for the scalar
curvature describing the global parametric intrinsic correlations.
By defining set of controller functions, we find explicitly that
the most general scalar curvature can be presented as
\begin{scriptsize}
\begin{eqnarray} \label{gsc}
R = - \frac{n}{2 D^2} \frac{\sum_{k=0}^5 (w_k r_k(1+f
S)^{kn}}{(S+b)(S-a)}
\end{eqnarray}
\end{scriptsize}
where the co-efficients $\{ r_i(a,b) \}$ appearing in the
numerator can be written as the following polynomials
\begin{scriptsize}
\begin{eqnarray}
r_0&=&S^{10}+10 S^9 b+45 S^8 b^2+120 S^7 b^3+210 S^6 b^4+252 b^5
S^5 \nonumber \\ &&
+210 b^6 S^4+210 b^7 S^3+45 b^8 S^2+10 b^9 S+b^{10}   \nonumber \\
r_1&=& -S^{10} +2(a- 4 b)S^9+(16 a b-28 b^2-a^2) S^8  \nonumber \\
&& +8(7 a b^2-7 b^3- a^2 b) S^7+14 (8 a b^3-5 b^4-2 a^2 b^2) S^6
\nonumber \\ && +28(5 a b^4-2 b^5-2 a^2 b^3) S^5+14(8 a b^5-2
b^6-5 a^2 b^4) S^4  \nonumber \\ && +8(7 a b^6- b^7-7 a^2
b^5)S^3+(16 a b^7- b^8-28 a^2 b^6) S^2  \nonumber \\ &&
+2ab (b^7-4 a b^6) S-a^2 b^8   \nonumber \\
r_2&=& -S^{10} +(4a-6 b) S^9 +(24 a b -15 b^2-6 a^2) S^8
\nonumber \\ && +(60 a b^2-36 a^2 b-20 b^3+4 a^3) S^7  \nonumber
\\ &&
 +(80 a b^3-15 b^4+24 a^3 b-90 a^2 b^2-a^4) S^6  \nonumber \\ &&
+(60 a^3 b^2-120 a^2 b^3+60 a b^4-6 a^4 b-6 b^5) S^5  \nonumber \\
&&
 +(80 a^3 b^3 -b^6 -90 a^2 b^4+24 a b^5-15 a^4 b^2)S^4  \nonumber \\ &&
+(60 a^3 b^4 +4 a b^6-20 a^4 b^3-36 a^2 b^5) S^3  \nonumber \\ &&
+(24 a^3 b^5-6 a^2 b^6-15 a^4 b^4) S^2  \nonumber \\ &&
+(4 a^3 b^6-6 a^4 b^5) S- a^4 b^6   \nonumber \\
r_3&=& S^{10} +(4 b-6 a) S^9 +(6 b^2+15 a^2-24 ab) S^8  \nonumber
\\ && +(4 b^3-20 a^3+60 a^2 b-36 a b^2) S^7  \nonumber \\ &&
 +( b^4-80 a^3 b+90 a^2 b^2+15 a^4-24 a b^3)S^6  \nonumber \\ &&
+(60 a^4 b-6 a^5-120 a^3 b^2+60 a^2 b^3-6 a b^4)S^5  \nonumber \\
&&
 +(a^6-24 a^5 b-80 a^3 b^3+90 a^4 b^2+15 a^2 b^4) S^4  \nonumber \\ &&
+(4 a^6 b-20 a^3 b^4+60 a^4 b^3-36 a^5 b^2) S^3  \nonumber \\ &&
+(6 a^6 b^2-24 a^5 b^3+15 a^4 b^4) S^2  \nonumber \\ &&
+(4 a^6 b^3-6 a^5 b^4) S+a^6 b^4   \nonumber \\
r_4&=& S^{10} +(2 b-8 a) S^9+( b^2-16 a b+28 a^2) S^8  \nonumber
\\ && +(56 a^2 b-56 a^3-8 a b^2 ) S^7+(70 a^4 +28 a^2 b^2-112 a^3
b)S^6  \nonumber \\ && +(140 a^4 b-56 a^3 b^2-56 a^5) S^5+(28 a^6
-112 a^5 b+70 a^4 b^2)S^4  \nonumber \\ && +(56 a^6 b-56 a^5 b^2-8
a^7) S^3+(a^8-16 a^7 b+28 a^6 b^2)S^2  \nonumber \\ &&
+(2 a^8 b-8 a^7 b^2)S+ a^8 b^2   \nonumber \\
r_5&=& S^{10} -10 a S^9 +45 a^2 S^8-120 a^3 S^7+210 a^4 S^6 -252
a^5 S^5   \nonumber \\ && +210 a^6 S^4 -120 a^7 S^3+45 a^8 S^2 -10
a^9 S+a^{10}
\end{eqnarray}
\end{scriptsize}
The weights $\{w_i\}$ occurring in the summation of the numerator
of Eqn.(\ref{gsc}) are given by the sequence
\begin{scriptsize}
\begin{eqnarray}
{w_i}:= \{-6(n-1),(9n-1),(10n+16),(8n-18),(16n+18), (n+1)\}
\end{eqnarray}
\end{scriptsize}
Furthermore, it turns out that $D(a,b)$ appearing in the
denominator of the scalar curvature, Eqn.(\ref{gsc}), is expressed
as the following function
\begin{scriptsize}
\begin{eqnarray}
D&=&(n-1)(S^4+4 S^3 b+6 S^2 b^2+4 S b^3+b^4)\nonumber \\ && +(1+f
S)^n (2n S^4+ 4n(b-a)S^3+ 2n(b^2-4ab+a^2)S^2+4nab(a-b)S+ 2na^2b^2)
\nonumber \\ && +(1+f S)^{2 n} (n+1)(S^4-4S^3 a+6 S^2 a^2-4 S
a^3+a^4)
\end{eqnarray}
\end{scriptsize}
Consequently, we may easily expose the associated important
conclusions for the specific considerations of the variable
mismatch factor controllers.
\begin{figure}
\hspace*{0.5cm}
\includegraphics[width=8.0cm,angle=0]{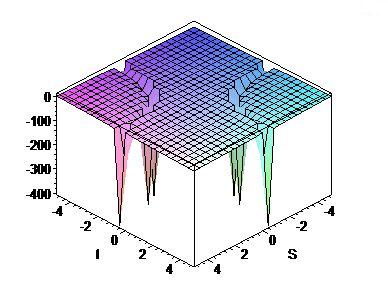}
\caption{The curvature scalar plotted as the function of a
variable mismatch factor $f$ and $S$ with given parameters $a$ and
$b$, describing the fluctuations in the controllers.}
\vspace*{-0.5cm}
\end{figure}
The global nature of phase transitions can be thus explored over
the range of parameters describing the controllers of interest.
For the limiting linear controllers corresponding to the values
$a=0, b=0$, the limiting intrinsic scalar curvature simplifies to
the following ratio of series
\begin{scriptsize}
\begin{eqnarray} \label{mcur}
R &=& \frac{\sum_{k=0}^5 t_k (1+f S)^{k n}} {((n+1) (1+f S)^{2
n}+2n (1+f S)^n+ n-1)^{2}}
\end{eqnarray}
\end{scriptsize}
where $t_i(n)$ are defined by the following sequence
\begin{scriptsize}
\begin{eqnarray}
t_i:= \{ 3n(n-1), \frac{9}{2}n(n-1), n(5n+8), -n(4n-9), -n(8n+9),
-\frac{1}{2}n(n+1)\}
\end{eqnarray}
\end{scriptsize}

\begin{figure}
\hspace*{0.5cm}
\includegraphics[width=8.0cm,angle=0]{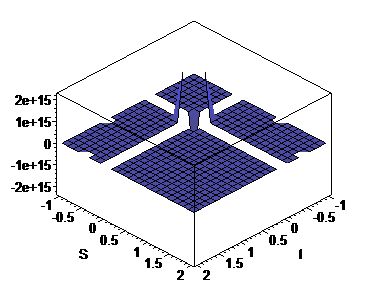}
\caption{The determinant of the metric tensor plotted as the
function of a variable mismatch factor $f$ and $S$ with given
parameters $a$ and $b$, describing the stability of the
controllers.} \vspace*{-0.5cm}
\end{figure}

\begin{figure}
\hspace*{0.5cm}
\includegraphics[width=8.0cm,angle=0]{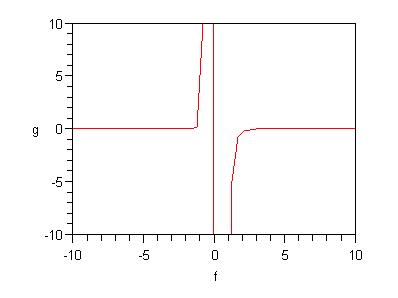}
\caption{The determinant of the metric tensor plotted as the
function of a variable mismatch factor $f$ with given parameters
$a$ and $b$, describing the fluctuations in the linear
controllers.} \vspace*{0.5cm}
\end{figure}
Eqn.(\ref{mcur}) shows that the global interactions exist for the
limiting linear values of the variable mismatch factor
controllers. This follows from the fact that the coefficients of
the scalar curvature Eqn.(\ref{mcur}), signifying the global
correlation volume of controller, remain non-zero in the linear
limit. In the limit of $n=1$, the Eqn.(\ref{mcur}) shows further
that the scalar curvature diverges on the $fS$-surface of the root
of $(2+ fS)^2$.

The graphical views of the curvature scalar of the variable
mismatch factor controllers are depicted in the two and three
dimensions. The intrinsic characterization offered in (i) Fig.[4]
is over the three dimensions and (ii) Fig.[7] is over the two
dimensions. This describes the precise global behavior of the
parametric fluctuations over the entire intrinsic manifold
$(M_3,g)$ for the limiting linear variable mismatch controllers
Fig.[2].

For the limiting linear controllers, it is worth mentioning that
the local pair correlations, as the components of the metric
tensor, have an expected behavior. The pure pair correlations
reduce as the Eqn.(\ref{limitingconstmismatchcorr}). The others
involving a variation of the mismatch factor reduce to the
following equations:
\begin{scriptsize}
\begin{eqnarray}
%g_{aa} &=& \frac{2}{S^2} \frac{(1+f S)^{2 n}+3 (1+f S)^n}{((1+f S)^n-1)^3}, \ \ % \nonumber \\
%g_{ab} = \frac{1}{S^2} \frac{(1+f S)^{2 n}+6 (1+f S)^n+1}{((1+f S)^n-1)^3} \nonumber \\
g_{af} &=& -n \frac{(1+f S)^{2 n-1}+3 (1+f S)^{n-1}}{((1+f S)^n-1)^3}, \ \ % \nonumber \\
%g_{bb} = \frac{2}{S^2} \frac{3 (1+f S)^n+1}{((1+f S)^n-1)^3} \nonumber \\
g_{bf} = -n \frac{(1+f S)^{2 n-1}+3 (1+f S)^{n-1}}{((1+f S)^n-1)^3} \nonumber \\
g_{ff} &=& S^2 n \frac{(1+f S)^{2 n-2} (n+1)+(1+f S)^{n-2}
(n-1)}{((1+f S)^n-1)^3}
\end{eqnarray}
\end{scriptsize}
In the case of the limiting linear controllers, we observe further
that the determinant of the metric tensor reduces to the following
expression:
\begin{scriptsize}
\begin{eqnarray}
Det(g) &=& - nS^{-2}(1+f S)^{n-2} ((1+f S)^n-1)^{-7} \nonumber \\
&& \bigg((n+1) (1+f S)^{3 n}+(3 n+1) (1+f S)^{2 n} \nonumber \\ &&
+(3 n-1)(1+f S)^n+ (n- 1) \bigg)
\end{eqnarray}
\end{scriptsize}
\begin{figure}
\hspace*{0.5cm}
\includegraphics[width=8.0cm,angle=0]{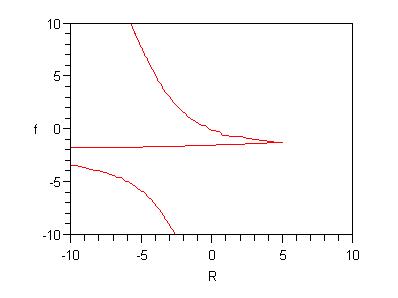}
\caption{The curvature scalar plotted as the function of a
variable mismatch factor $f$ with given parameters $a$ and $b$,
describing the fluctuations in the linear controllers.}
\vspace*{0.5cm}
\end{figure}
It is important to notice that the global stability of the
controllers may be determined by observing the sign of the
determinant of the metric tensor. For $n=1$, we find that the
determinant of the metric tensor reduces to the following
expression:
\begin{scriptsize}
\begin{eqnarray}
Det(g)= -\frac{2}{S^2}
\frac{(1+fS)^2+2(1+fS)^3+(1+fS)^4}{(1+fS)^{2}((1+fS)-1)^{7}}
\end{eqnarray}
\end{scriptsize}
The behavior of the determinant of the metric tensor shows that
the variable mismatch factor controller becomes unstable for the
specific values $|f| \le 1.2$ of the associated mismatch factor.
For the general value the $S$, the nature of the determinant of
the metric tensor is depicted in the Fig.[5]. For $S=1$ and $n=1$,
the corresponding surface nature of the determinant of the metric
tensor is depicted in the Fig.[6].

In contrast to the constant mismatch factor controllers, we notice
in the present section that the determinant of the metric tensor
reduces to the cusp form in the regime of $f \rightarrow 0$. In
this domain of the mismatch factor, we observe that the variable
mismatch factor controllers are relatively less stable than the
constant mismatch factor controllers. The corresponding surface
behavior of the scalar curvature is depicted in the Fig.[7]. This
describes the global phase properties of the variable mismatch
factor controllers.

It is shown that the non-zero value of intrinsic scalar curvature
further demonstrate the existence of a finite correlation volume.
Specifically, the phase stability of typical controllers with a
variable mismatch factor can thus be easily determined by
analyzing the nature of the scalar curvature in the domain of
interest. This has been depicted in the Figs.[4, 7], in which we
show the global nature of the parametric correlations.

The further observation of the Fig.[4] shows that the variable
mismatch factor controller systems have no phase transitions on
the parametric manifold $(M_3,g)$. Subsequently, the global nature
of variable mismatch factor controllers is well explicable against
the local fluctuations of the model parameters $\{a, b\}$ and
mismatch factor $f$ of the controller.
\section{Conclusion}
The intrinsic geometric design of controllers is offered under the
fluctuations of the model parameters and mismatch factor. Such
fluctuations are expected to arise due to non-zero heating
effects, chemical reactions and possible conventional corruptions
associated with the controller under the application. The
intrinsic geometric method is used to improve the structure of the
bounded parametric controller id thus designed. It is pictorially
presented for the limiting polynomial approach corresponding to
the limiting linear parametrization. The stability analysis thus
introduced is most generic for the fluctuations of the parameter
and the mismatch factor the controllers.

The present analysis is well suited for practical applications.
The intrinsic geometric design procedure is presented for the
controllers with a (i) constant and (ii) variable mismatch factor.
In this concern, the Fig.(3) and Fig.(6) show the respective
determinants of the metric tensor for the constant and variable
mismatch factor controllers. These figures illustrate that the
typical instability appears as (i) a throat for the constant
mismatch factor controllers and (ii) a cusp for the variable
mismatch factor controllers. Subsequently, a straightforward
comparison may be made between the stability properties of the
constant and the variable mismatch factor controllers. In the
first case, it turns out that the associated controllers
correspond to a non-interacting system, while in the second case
such a controller configuration corresponds to an interacting
system. This follows from the fact that the manifold of parameters
is flat in the first case, while it becomes curved for the
variable mismatch factor controllers.

In the limit of $f \rightarrow 0$, we have shown in the first case
that the determinant of the metric tensor acquires a throat,
whereas the determinant of the metric tensor of the variable
mismatch factor controllers acquires a cusp in this limit. Thus,
the present investigation predicts that the controller systems
with the constant mismatch factor are relatively more stable and
better-behaved than those with the variable mismatch factor. In
addition, our model is well suited for the robust controllers.
Such controllers are very popular now a days, because of their
high performance and low maintenance needs. From the commercial
viewpoints, such a robust controller is very lucrative. It is
worth mentioning that the use of the intrinsic geometric principle
is rapidly growing in robust controller design in recent years.

Based on the definition of the controller, the intrinsic stability
analysis remains compatible for parametrically stable designs of
the controller and their modern appliances. The present analysis
thus provides the intrinsic geometric front to the stability
analysis of existing controllers and their possible future
generations. It may be also used, in order to model in a suitable
fashion the un-modeled part of the plant and the bounded
disturbances. Finally, it is envisaged that our analysis offers
perspective stability grounds, when applied to the electrical
plants. It is expected further that the present investigation
would be an important factor in an appropriate design of the
safety guards, which can work as the indicators under fluctuations
of the parameters, mismatch factor and the other possible
components.
\section*{Acknowledgment}
%
%The work of SB is supported by the European Research Council grant n.~226455,
%\textit{``SUPERSYMMETRY, QUANTUM GRAVITY AND GAUGE FIELDS (SUPERFIELDS)''}.
B. N. T. thanks Prof. V. Ravishankar for the support and
encouragement. The work of BNT has been supported by the
postdoctoral research fellowship of \textit{``INFN-Laboratori
Nazionali di Frascati, Roma, Italy''}.

\end{document}